\documentclass[a4paper,aps,prl,twocolumn,floatfix,showpacs,superscriptaddress,footnoteinbib]{revtex4-1}
\usepackage{blindtext}
\usepackage{graphics}
\usepackage{epsfig}
\usepackage{bbm}
\usepackage{times}
\usepackage{xcolor}   
\usepackage{amsfonts}
\usepackage{amsmath}
\usepackage{amssymb}
\usepackage{amsthm}

\usepackage{wasysym}
\usepackage{bm} 

\usepackage[colorlinks=true,linkcolor=blue,citecolor=blue,urlcolor=blue]{hyperref}
\usepackage{url}

\begin{document}

\title{Anomalous Landau Levels in Inhomogeneous Fluxes and Emergent Supersymmetry}

\author{Soujanya Datta}
\affiliation{Theory Division, Saha Institute of Nuclear Physics, 1/AF Bidhannagar, Kolkata 700064, India}
\affiliation{Homi Bhabha National Institute, Training School Complex, Anushaktinagar, Mumbai 400094, India}
\author{Krishanu Roychowdhury}
\affiliation{Theory Division, Saha Institute of Nuclear Physics,
1/AF Bidhannagar, Kolkata 700064, India}
\affiliation{Homi Bhabha National Institute, Training School Complex, Anushaktinagar, Mumbai 400094, India}
\affiliation{Max-Planck-Institut f\"{u}r Physik komplexer Systeme,
N\"{o}thnitzer Strasse 38, 01187 Dresden, Germany}

\begin{abstract}
Two-dimensional (2D) systems in magnetic fields host rich physics, most notably the quantum Hall effect arising from Landau level (LL) quantization. In a broad class of 2D models, flat bands with topologically nontrivial band degeneracies give rise to anomalous LL quantization under homogeneous fields. Ascribed to the underlying quantum geometry, these are classified as {\it singular flat bands} (SFBs), exhibiting unusual wavefunction localization, and anomalous quantization of LLs. We investigate the response of gapless SFBs to inhomogeneous fluxes, bridging continuum and lattice descriptions. Our analysis reveals a mechanism to controllably manipulate the anomalous LLs via flux inhomogeneity. We further uncover an emergent supersymmetry (SUSY) in the parameter space where the tower of ALLs collapses to zero energy, rendering a {\it lattice analog of the Aharonov-Casher theorem} on degenerate zero modes in perpendicular fluxes, with wavefunction localization partly similar to Aharonov–Bohm caging. With the addition of strong correlations, these findings will have implications for realizing exotic topological and charge-ordered phases.
\end{abstract}

\maketitle



\noindent
{\color{blue}{\it Introduction.}} Classification of quantum matter into categories of metals and insulators fundamentally relies on the details of the electronic band structures \cite{bloch1929quantenmechanik, wilson1931theory, kohn1964theory}. Key features such as band crossings, topological twists in isolated bands, and band curvature (linked to {\it e.g.}, inflection points, effective mass tensor, group velocities) play crucial roles in determining the behavior of the quasiparticles in those systems across relativistic and nonrelativistic regimes \cite{yu2024quantum}. When protected by symmetries, these structural elements underpin a wide spectrum of exotic phenomena, culminating, among others, in the discovery of topological quantum matter (TQM) \cite{kane2005quantum, bernevig2006quantum, fu2007topological, roy2009topological, moore2010birth}.

In the wake of TQM, another characteristic that has emerged as a central concept is quantum geometry \cite{hwang2021geometric,torma2023essay}. Encoded in the quantum metric and Berry curvature \cite{shapere1989geometric}, it admits symmetry regulations with measurable consequences in diverse physical observables. Yet, symmetries are not always essential for stabilizing topological features. A prototypical example is the 2D quantum Hall insulator, where dispersive bands split into discrete flat Landau levels (LLs) when immersed in a homogeneous perpendicular magnetic field that breaks the time-reversal symmetry \cite{Landau1930, klitzing1980new, Thouless1982}. In both nonrelativistic and relativistic settings (Dirac/Weyl type \cite{McClure1956}), such LL structures can be accurately predicted by Onsager’s semiclassical quantization \cite{onsager1952interpretation}, which derives from Einstein-Brillouin-Keller condition \cite{Einstein1917, Brillouin1926, Keller1958}: For a set of generalized coordinates ${\bf q}$ and conjugate momenta ${\bf p}$, the closed action $\oint {\bf p}\cdot {\rm d}{\bf q}$ is quantized up to topological corrections, proportional to the strength of the magnetic field. 

Remarkably, the quantum Hall paradigm extends into a broader class of systems known as anomalous quantum Hall \cite{haldane1988model} insulators (sometimes referred to as Chern insulators \cite{regnault2011fractional, thonhauser2006insulator}), where a Hall-like response arises even in the absence of an external magnetic field. This behavior stems from the topological nature of the underlying band structures generated by staggered flux configurations  \cite{haldane1988model, xu2015intrinsic, lan2023flat} and has encouraged significant experimental exploration \cite{zhao2024realization}. Certain spin-orbit-coupled ferromagnetic kagome materials \cite{xu2015intrinsic} furnish a natural setting for realizing such inhomogeneous fluxes, maintaining zero net flux through the unit cell. More versatile platforms include the Van der Waal heterostructures (where gauge fields have spatial modulation over much larger length scales than typical lattice spacings, and the enlarged unit cell can accommodate net flux \cite{Phong2022BoundaryModes, PhongMele2025ZeroModes}), which offer exquisite tunability and mechanistic control over the band engineering, housing a rich palette of unconventional phases, accessible through transport and spectroscopic probes \cite{bistritzer2011moire, cao2018correlated, cao2018unconventional, andrei2021marvels}. Additionally, these materials have garnered tremendous attention, being a prime host of nearly flat bands, responsible for dominant electronic correlations. As such, quantum materials with flat bands are of fundamental importance in condensed matter physics, featuring topological responses as well as exotic ordering phenomena \cite{lin2018flatbands, huang2024non, kang2020dirac, li2018realization, lee2020stable}.

A variety of lattice models, such as the kagome lattice, naturally fosters perfectly flat bands, either isolated in the spectrum or intersecting some of the dispersive bands at discrete points, lines, or surfaces \cite{Mielke1991, Tasaki1992, Bergman2008, Leykam2018}. The simplest yet illustrative case involves a two-band model (which can be regarded as an effective low-energy description of some tight-binding models) in which a quadratic band touches a flat band at one point (extendable to multiple flat bands). These are exotic band degeneracies guarded either by crystal symmetries \cite{mizoguchi2021qbt}, textured fluxes (this work), or a recently proposed lattice supersymmetry (SUSY) \cite{roychowdhury2024supersymmetry}. The singularities of flat bands can be characterized through the quantum geometric tensor \cite{rhim2020quantum}, while the compact localization of single-particle states bears striking differences between singular and nonsingular flat bands \cite{rhim2019classification}.

When such models are subjected to a homogeneous perpendicular magnetic field, conventional LLs are not observed due to the breakdown of Onsager's condition. Instead, for singular flat bands (SFBs), {\it anomalous Landau levels} (ALLs) appear below the flat band energy, originating from a repulsive coupling between the LLs of the dispersive band and the flat band \cite{rhim2020quantum}. This coupling is governed by the finite quantum metric at the gap-closing singularity. 
Our motivation is to develop a theory of ALLs under inhomogeneous, periodically textured fluxes and unveil a mechanism for tuning their bandwidths with a single modulation parameter set by the flux configuration. This analysis clarifies the conditions under which such LL anomalies occur in flat-band systems and highlights how they may be manipulated in ways accessible to transport experiments. Curiously, en route, we discover an emergent SUSY in a special region of the phase diagram when depicted in terms of the flux modulation parameter, where all ALLs collapse to form a macroscopic degenerate manifold of zero-energy states, determined solely by the flux strength, also evinced by the tight-binding spectrum. A systematic exposition of the findings follows.

\noindent
{\color{blue}{\it The continuum model.}} We begin with a two-band model that shows a flat band at zero energy and a quadratic band touching it at ${\bf k}=0$. The Hamiltonian, in the minimal description, can be expressed in terms of four parameters $a_1,a_2,a_3,a_4$ as
\begin{align}\label{Ham1}
 {\cal H}({\bf k}) = \sum_\mu d_\mu({\bf k}) \sigma_\mu\,,
\end{align}
with
\begin{align}\label{params1}
  d_0({\bf k}) &= \frac{1}{2}\left[ a_1^2 k_x^2 + \left(a_2^2+a_3^2+a_4^2  \right) k_y^2 + 2 a_1 a_2 k_x k_y    \right], \nonumber \\
  d_1({\bf k}) &= a_3 a_4 k_y^2 ~,~~  d_2({\bf k}) = a_2 a_4 k_y^2 + a_1 a_4 k_x k_y, \nonumber \\
  d_3({\bf k}) &= \frac{1}{2}\left[ a_1^2 k_x^2 + \left(a_2^2+a_3^2-a_4^2  \right) k_y^2 + 2 a_1 a_2 k_x k_y    \right], \nonumber
\end{align}
and $\sigma_\mu$, the Pauli matrices. The condition ${\rm Det}[{\cal H}({\bf k})]=0$ pins the flat band at $E=0$. The band degeneracy at ${\bf k}=0$ defines a gapless SFB protected by a finite Hilbert-Schmidt quantum distance \cite{provost1980riemannian, buvzek1996quantum, dodonov2000hilbert} $d^2=1- |\langle \psi({\bf k}_1)|\psi({\bf k}_2) \rangle|^2$ between two states on the flat band located symmetrically about the degeneracy point along a line -- for an SFB, this attains a maximum $d^2_{\rm max}=a_4^2/(a_3^2+a_4^2)$ as $|{\bf k}_{1}-{\bf k}_{2}| \rightarrow 0$. This characteristic facilitates a broad classification of the SFBs into {\it weak} and {\it strong} categories -- depending on the limit $d^2_{\rm max}$ approaches with ${\bf k}_1 \rightarrow {\bf k}_2$, a weak SFB corresponds to $d^2_{\rm max} \ll 1$ (for $a_3 \gg a_4$) while for a strong one, $d^2_{\rm max}\rightarrow 1$ (for $a_3 \ll a_4$). For a non-SFB, $d^2_{\rm max} \rightarrow 0$ as $|{\bf k}_{1}-{\bf k}_{2}| \rightarrow 0$ \cite{rhim2019classification}.    

In the presence of a homogeneous perpendicular magnetic field (of strength $B$), minimal coupling leads to the quantization of energy, forming discrete LLs. For dispersive bands, this yields the familiar scalings \( E_n \propto \pm \sqrt{n} \) for relativistic and \( E_n \propto n \) for nonrelativistic cases, as captured by Onsager’s semiclassical quantization relation \cite{onsager1952interpretation}. However, this framework breaks down for flat bands due to an ill-defined Fermi surface area. In particular, for SFBs in a homogeneous magnetic field, the LLs exhibit an anomalous scaling \( E_n \propto (-1/n) \) below the flat band energy with a finite bandwidth determined by the combination $d^2_{\rm max} B$. It has been argued that as long as a nonzero quantum distance $d^2_{\rm max}$ persists to characterize the band degeneracy, such an anomaly will be observed even for a weak SFB. This triggers a natural question: Does such anomalous quantization appear in inhomogeneous magnetic fields, and in lattice realizations? Compared to the homogeneous case, the inhomogeneous setting reveals a richer and more nuanced behavior -- central to the key findings of this work.

We proceed by sketching the derivation of the ALL quantization for the homogeneous case. Fixing the gauge that breaks translation symmetry along the \( x \)-direction, hence \( k_x \rightarrow -i\partial_x \), \( k_y \rightarrow k + x/l^2 \), $l$ being the magnetic length \( l = \sqrt{\hbar/eB} \), one obtains a real-space representation of the Hamiltonian matrix in Eq.~\ref{Ham1}, denoted \( \mathcal{H} \). This naturally entails the noncommutative algebra $[k_x,k_y]=-i/l^2$. We seek normalizable solutions to the eigenvalue equation \( \mathcal{H}\Psi = E\Psi \), for energies close to the flat band energy $E=0$. Adopting the ansatz \( \Psi(x, y) = e^{i k y} \begin{bmatrix} \phi_A(x) & \phi_B(x) \end{bmatrix}^{\rm T} \), to leading order in $E$, we arrive at the following differential equation to solve \cite{suppl} 
\begin{equation}\label{diffeq1}
    \nu^2\partial_\nu^2 \phi_A - \frac{\nu}{2} \partial_\nu \phi_A - [P + Q\nu + R  \nu^2]\phi_A = 0\,,
\end{equation}
where $\nu = \left(x + kl^2 \right)^2$, $P={7}/{16}$, $Q=3a_4^2/\left(16 El^4\right) + a_3/\left(2a_1l^2\right)$, $R=\left(a_2^2 + a_{3}^2 + a_{4}^2\right)/\left(4 a_1^2 l^4\right)$. The square-integrability condition on the confluent hypergeometric solutions determines the ALL energy quantization \cite{suppl}
\begin{align}\label{enladder1}
    E_n = - \frac{1}{8} \left|\frac{3a_1a_4^2/l^2}{(2n+3)\sqrt{a_2^2 + a_{3}^2 + a_{4}^2}+a_3} \right|,
\end{align}
with degeneracy fixed by the allowed $k$-values, as in conventional LLs.
This expression identifies the necessary conditions for the existence of ALLs: both \(a_1\) and \(a_4\) must be nonzero. In particular, the quantum distance vanishes as \(a_4 \to 0\), separating weak from strong SFBs. For weak SFBs (with \(a_3=a_1,\, a_4 \approx a_1 d_{\rm max}\)) and setting \(a_2=0\) for simplicity, the LL energies scale with the quantum distance as \(E_n \approx -(3a_1^2 d_{\rm max}^2/16l^2)(1/n)\) for large \(n\). By contrast, in strong SFBs (\(a_3=0\)), they become independent of \(d_{\rm max}^2\), yielding \(E_n \approx -(3a_1 a_4/16l^2)(1/n)\). These results underscore the pivotal role of quantum geometry in the formation of ALLs, establishing a nonzero quantum distance as, in general, a necessary condition for their emergence \cite{rhim2020quantum}. However, as we show below, inhomogeneous, periodically modulated flux configurations (not necessarily staggered) can provide a notable exception, unveiling a richer landscape. To address this, we first turn to lattice model calculations and subsequently reformulate the findings within the continuum framework.

An alternative derivation of the ALL quantization follows by casting the Hamiltonian \( \mathcal{H} \) in terms of ladder operators \cite{rhim2020quantum}, 
\(k_x = (a + a^\dagger)/\sqrt{2}l\), \(k_y = i(a - a^\dagger)/\sqrt{2}l\), which preserve the commutation algebra. 
Expressed in the number basis of \(\hat{n} = a^\dagger a\), the Hamiltonian acquires a block structure: a block labeled by the eigenvalue index \(n\) couples only to its neighbors 
\((n\pm 2)\) through terms such as \(a^2\) and \(a^{\dagger 2}\), reflecting the quadratic dependence of \({\cal H}({\bf k})\) on \(k_x, k_y\). This sufficiently captures the essential spectral features of the ALLs and their dependence on the quantum geometry. 

\begin{figure*}
 \centering
  \includegraphics[width=2.07\columnwidth]{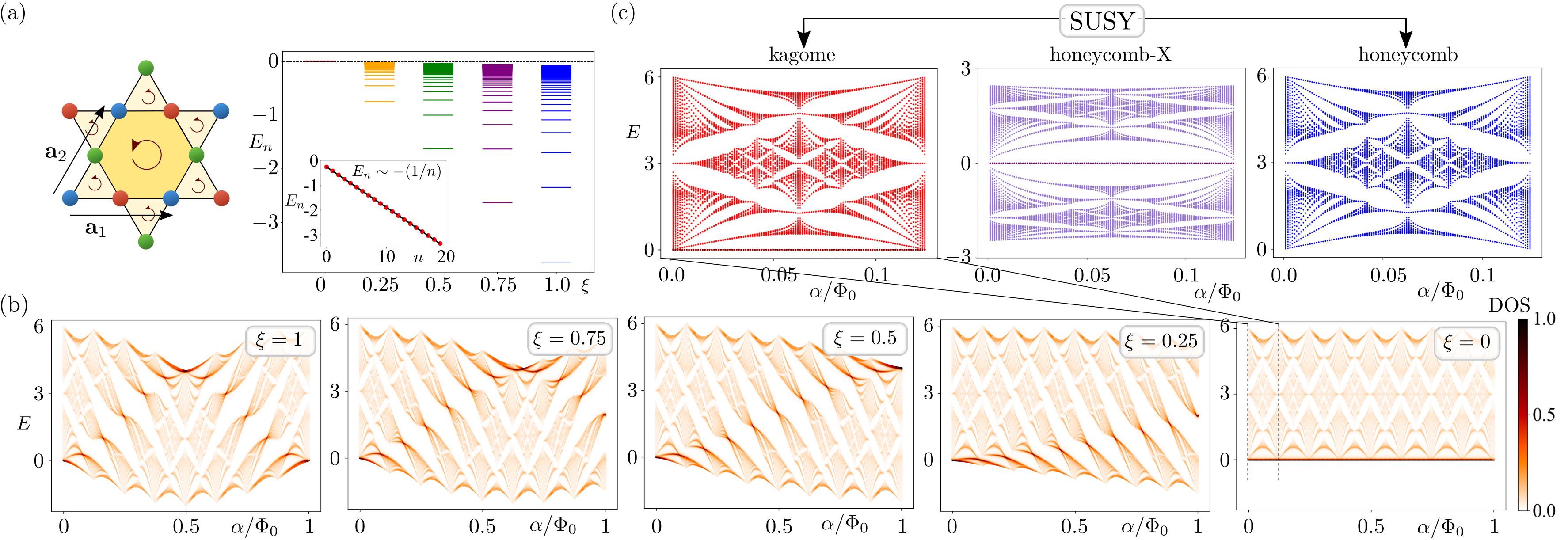}
  \caption{(a) Left: The kagome lattice SUC specified by the lattice vectors ${\bf a}_{1,2}$, the honeycomb plaquette is pierced by flux $\Phi_{\hexagon}=(8-2\xi)\alpha$, while each triangular plaquette by $\Phi_{\Delta}=\xi\alpha$, making the total flux content of the SUC to be $8\alpha$. Right: kagome ALL spectra as a function of the modulation parameter $\xi$ display a gradual squeezing from the uniform limit at $\xi=1$ to the SUSY limit $\xi=0$, computed for $p=1, q=113$. The inset shows the anomalous energy quantization $E_n\sim -1/n$ at $\xi=1$. (b) The nature of modulation set by $\xi$ can be best understood by inspecting the HRBS fractal revealed through the normalized DOS: their absence indicates the gaps while a dense population marks the degeneracy, most prominent at the SUSY point ($\xi=0$). (c) At the SUSY point, the HRBS repeats at $\alpha/\Phi_0=1/8$ as argued in the text. The zoomed-in version is shown in red. The lattice SUSY explained in the text implies the existence of its partner with an identical HRBS but without the zero modes (honeycomb lattice, blue), connected by the chiral SUSY charge at the middle (honeycomb-X lattice, violet). The kagome sector shares the zero modes with the charge lattice, the latter being the square root of the former.}
 \label{fig1}
\end{figure*}

\noindent
{\color{blue}{\it The lattice model.}} To illustrate the emergence of ALLs in a lattice setting, we consider spinless fermions hopping on a kagome lattice described by the nearest-neighbor (nn) tight-binding Hamiltonian ${\cal H}_{\rm TB} = t \sum_{\langle ij \rangle} c_i^\dagger c_j + {\rm h.c.} + \mu \sum_i c_i^\dagger c_i$, where $c_i^\dagger$ creates a fermion on site $i$. For positive hopping amplitude ($t>0$), the lowest band is flat, touching a quadratically dispersing band at the $\Gamma$ point (${\bf k}=0$). A perpendicular magnetic field modifies the hopping amplitudes via Peierls substitution, $t \rightarrow t e^{i\phi_{ij}}$, where the phases $\phi_{ij}$ are assigned to the nn bonds such that their circulation around a plaquette equals the magnetic flux threading that plaquette.

The kagome structural unit cell (SUC), specified by the lattice vectors $\bf{a}_{1,2}$, consists of one hexagon and two triangles [Fig.~\ref{fig1} (a)], with fluxes $\alpha$ (triangle) and $6\alpha$ (hexagon), giving total SUC flux $\Phi_{\rm SUC}=8\alpha$ (mod $\Phi_0$). For rational flux $\Phi_{\rm SUC}/\Phi_0=p/q$, we have $\alpha/\Phi_0=p/8q$, $p$ varying from $1,\dots,(8q-1)$. The magnetic unit cell (MUC) is then $q$-fold enlarged along $\bf{a}_1$ (in the gauge used for the continuum analysis).
The spectrum obtained by diagonalizing ${\cal H}_{\rm TB}$ is periodic in \( \alpha \), repeating whenever \( \alpha \) is an integer multiple of \( \Phi_0 \). The flat band in the lattice model can be pinned at energy $E=0$ by fixing the chemical potential to $\mu=2$. Collating the spectrum at the $\Gamma$ point for different values of $\alpha/\Phi_0\in (0,1)$, a butterfly-like fractal pattern emerges, known as the Hofstadter-Rammal butterfly spectrum (HRBS) \cite{hofstadter1976energy, li2011tight}, symmetric about $\alpha/\Phi_0=1/2$. The LLs appearing below $E=0$ in this spectrum at $\alpha/\Phi_0$ close to $0$ exhibit the characteristic anomalous scaling $E_n \propto (-1/n)$, as evident from Fig.~\ref{fig1} (a). With the homogeneous case understood, we now turn to the more intriguing question of how ALLs evolve under inhomogeneous flux modulations.

\noindent 
{\color{blue}{\it ALLs under inhomogeneous fluxes.}} We implement flux inhomogeneity in the lattice model, keeping the total flux of the kagome SUC fixed, through a modulation parameter $\xi$, such that each triangular plaquette carries a flux $\Phi_{\Delta} = \xi \alpha$, while the hexagon hosts $\Phi_{\hexagon} = (8-2\xi)\alpha$, with the gauge choice unchanged. The homogeneous case is recovered at $\xi=1$, for which the anomalous scaling has already been discussed.
We are now equipped to explore a potentially uncharted territory by tuning away from the homogeneous limit toward $\xi=0$ (a special point for reasons to be clear later) and beyond, while keeping the size of the MUC unchanged. For clarity, we restrict $\xi$ to the interval $[0,1]$ and assign the complex phases to the hopping amplitudes using the following convention (see Fig. S1 in \cite{suppl}): indexing the SUCs within a MUC by an integer $l \in [1,q]$, the nonzero phases for the $l$-th SUC are $\phi^{(1)}_l = 8l\alpha, \phi^{(2)}_l = \xi \alpha, \phi^{(3)}_l = [\xi - 8(l+1)]\alpha$. With this assignment, each triangular plaquette carries flux $\Phi_{\Delta} = \phi^{(2)}_l = \xi \alpha$, while each hexagon carries flux $\Phi_{\hexagon} = -\phi^{(1)}_l - \phi^{(2)}_l - \phi^{(3)}_l = (8 - 2\xi)\alpha$, independent of $l$. The spectrum of this model is obtained by diagonalizing ${\cal H}_{\rm TB}$ for one MUC, which, for a fixed $(p,q)$, corresponds to a $3q \times 3q$ matrix. Stacking the eigenvalues at ${\bf k}=0$ as $\alpha/\Phi_0$ varies reveals how flux inhomogeneity reshapes the kagome HRBS and the spectral fingerprints of the ALLs.

For all values of $\xi$, the fractal structure of the kagome HRBS is embellished with major and minor gaps, where the closing of a major gap signals a topological phase transition. To highlight these features, the evolution of the HRBS with flux inhomogeneity is presented via the (normalized) density of states (DOS) in Fig.~\ref{fig1} (b) \cite{suppl}. A complementary probe is the edge locality marker, which confirms the presence of edge states inside the major gaps \cite{suppl}. For intermediate values $\xi \in (0,1)$, the spectrum loses its periodicity over $\alpha = \Phi_0$ and the symmetry about $\alpha/\Phi_0 = 1/2$, recovering them only at $\xi = 0$. A similar asymmetry arises when the kagome lattice is irradiated with circularly polarized light \cite{du2018floquet}; however, unlike the bandwidth squeezing observed in that case, here the overall bandwidth remains approximately unchanged, with only modest distortion of the states near $E=0$ and a more limited redistribution of the spectral weight in the remaining bands. The ALLs occupy a narrow region below $E=0$ for $\alpha/\Phi_0 \ll 1$ (small $p/q$), with $(q-p)$ levels exhibiting the energy quantization $E_n \approx -\epsilon_0/n$.

A close inspection of the HRBS evolution with the inhomogeneity parameter $\xi$ reveals a key observation: the width of the ALL tower, $\epsilon_0$, decreases nearly linearly as $\xi$ is tuned from $1$ to $0$, with only minor nonlinear corrections. Lattice calculations yield $\epsilon_0 \sim -(\alpha/\Phi_0)(2.44 \, \xi + 1.05 \, \xi^2)$, while the ladder-operator approach, discussed previously, predicts $\epsilon_0 l^2 \sim -( 0.14 \, \xi + 0.04 \, \xi^2)$ \cite{suppl}.

\noindent
{\color{blue}{\it An emergent SUSY.}} The collapse of the ALLs at $E=0$ for the inhomogeneous flux configuration at $\xi=0$ (where flux threads only kagome hexagons), regardless of the magnetic field strength $\alpha/\Phi_0$, is a direct consequence of the positive semi-definiteness of the kagome MUC Hamiltonian, denoted $H^{(1)}({\bf k})$. For fixed magnetic flux $p/q$, the index theorem guarantees $q$ zero modes, one per SUC. This structure admits a lattice SUSY factorization $H^{(1)}({\bf k}) = R^\dagger({\bf k}) R({\bf k})$, introducing a partner Hamiltonian $H^{(2)}({\bf k}) = R({\bf k}) R^\dagger({\bf k})$ (or {\it vice-versa}), with finite-energy eigenvalues shared and zero-mode degeneracies differing by the Witten index $\nu$. The SUSY partner of the kagome lattice at $\xi=0$ turns out to be a honeycomb lattice subjected to homogeneous fluxes. This correspondence follows from the square root of $H^{(1)}({\bf k})$ yielding a tight-binding model on the honeycomb-X lattice, but in homogeneous fluxes like the honeycomb model. This is a bipartite lattice, with one sublattice forming a kagome network and the other, a honeycomb network. The honeycomb-X MUC Hamiltonian can be written in the chiral form $\begin{bmatrix}
  & R({\bf k}) \\
  R^\dagger({\bf k}) &
\end{bmatrix}$; hence, its square produces the two decoupled blocks $H^{(1,2)}({\bf k})$. This is precisely the Hermitian charge operator that generates the lattice SUSY \cite{roychowdhury2024supersymmetry}. While the details of the underlying algebra can be found in Ref.~\cite{roychowdhury2024supersymmetry}, we note the Witten index of this SUSY theory is \(\nu = q\), ensuring $q$ zero modes for the kagome MUC sector, which are absent in the HRBS on the honeycomb side; the rest looks identical. The square root of either sector's HRBS reproduces that of the honeycomb-X model, a direct manifestation of the above lattice SUSY. Attributed to the chiral symmetry, the spectrum is symmetric about $E=0$, exhibiting a degenerate manifold of zero modes. Furthermore, the spectral structure of all HRBSs (related by squaring or square-rooting) is periodic in $\alpha/\Phi_0$ with period $1/8$, consistent with the triangular plaquettes on the kagome lattice (each occupying $1/8$ of the total area of a SUC) being flux-free. This periodicity of the kagome HRBS at $\xi=0$ is inherited by all SUSY-linked models, completing the lattice triptych in Fig.~\ref{fig1} (c). These results realize a {\it lattice analog of the Aharonov–Casher (AC) theorem}: originally formulated in a continuum setting, the theorem predicts a manifold of zero modes of dimension $N$ to appear when the total flux satisfies $\lfloor\Phi_{\rm tot}/\Phi_0\rfloor=N$. In our lattice model, the macroscopic degeneracy of kagome ALLs at $E=0$ is likewise set solely by the external magnetic flux, but is rooted in an emergent lattice SUSY that depends sensitively on the flux distribution within the MUC.  Unlike the continuum case, the arrangement of flux among the simplices of the MUC controls whether and how the collapse to $E=0$ occurs. We further note that the honeycomb-X partner also hosts a flat band at $E=0$ in zero field, but chiral symmetry prevents the formation of ALLs under an applied magnetic field.

\noindent
{\color{blue}{\it Wavefunction localization.}} At the SUSY point, the lowest band becomes flat, enabling compact localization of the single-particle states in a magnetic field. While reminiscent of Aharonov–Bohm caging \cite{vidal1998ab} due to destructive interference of wavefunctions amplitudes, the effect here, however, corresponds to an inhomogeneous distribution of fluxes in the SUC, instead of a uniform specific value of the flux for all plaquettes, and hence, restricted to the lowest band only. For such a state with amplitudes $\psi_G({\bf R})$ and $\psi_R({\bf R})$ respectively on the green and red site of the SUC at ${\bf R}$ (Fig.~\ref{fig1}), a leakage from the hexagon is given by ${\cal A}=t\left[\psi_G({\bf R})+\psi_R({\bf R})\right]$; the compact localization ensues from ${\cal A}=0$ \cite{bergman2008band}. In the presence of the fluxes, our gauge assignments on the bonds modify the leakage to ${\cal A}=t\psi_G({\bf R})\left[1-e^{i\xi\alpha/\Phi_0}\right] \propto \xi \alpha $ for small $\xi, \alpha$. This simple argument immediately makes it clear why flat bands reappear in the spectrum whenever the flux through the kagome triangles is zero, or integer multiples of $\Phi_0$. In the latter, the corresponding $\xi$ values fall beyond $[0,1]$. 

\noindent
{\color{blue}{\it The continuum model at the SUSY point.}} The emergent SUSY discussed above offers a notable advantage. Although deriving an effective continuum description consistent with lattice results for arbitrary inhomogeneous fluxes is formidable, our SUSY framework enables such a construction for the kagome lattice under a wide class of inhomogeneous flux configurations near the SUSY point. To understand this, we carry out the SUSY algebra by taking the continuum limit of the charge operator $\mathcal{Q}$ for the honeycomb-X lattice, obtained by expanding its tight-binding Hamiltonian around ${\bf k}=0$. Upon projecting out the isolated bands away from $E=0$, we arrive at an effective three-band model whose spectrum exhibits a Dirac cone interlaced with a flat band at $E=0$ pinned by chiral symmetry. In fact, a more generic four-parameter description applies here by expressing the SUSY charge $\mathcal{Q}({\bf k})$ in terms of the Gell-Mann matrices $\lambda$'s as
\begin{align}\label{charge}
  {\cal Q}({\bf k}) = a_3 k_y \lambda_4 + (a_1 k_x + a_2 k_y)\lambda_5 + a_4 k_y \lambda_6\,,
\end{align}
corresponding to an anisotropic honeycomb-X lattice model. Exposing to a homogeneous magnetic field imposes substitutions \( k_x \rightarrow -i\partial_x \equiv q_x \), \( k_y \rightarrow k + x/l^2 \equiv q_y \) such that $[q_x,q_y]=-i/l^2$, giving rise to a modified SUSY charge, which can be squared to complete the SUSY algebra. This leads to a two-band Hamiltonian for the kagome lattice in a specific inhomogeneous flux setting which, consistent with the corresponding lattice model, reads
\begin{align}\label{Ham2}
 {\cal G}({\bf q}) = \sum_\mu b_\mu({\bf q}) \sigma_\mu\,,
\end{align}
with
\begin{align}\label{Ham2a}
  b_0({\bf q}) &= d_0({\bf q}) + a_1(ia_2-a_3)/(2l^2), \nonumber \\
  b_1({\bf q}) &= d_1({\bf q}) - a_1a_4/(2l^2)  ~,~~  b_2({\bf q}) = d_2({\bf q}) + ia_1a_4/l^2, \nonumber \\
  b_3({\bf q}) &= d_3({\bf q}) + a_1(ia_2-a_3)/(2l^2), \nonumber
\end{align}
with the $\bf d$ vector defined in Eq.~\ref{Ham1}. The ordering of $q_x$ and $q_y$ is crucial to ensure $\mathbf{b}$ is real, unlike the $k_x,k_y$ ordering in Eq.~\ref{Ham1}. Solving the corresponding differential equations obtained from the Schr\"{o}dinger equation $[{\cal G}(x)-E]\Phi=0$ confirms nontrivial (normalizable) solutions at $E=0$, which serves as the lower bound of the spectrum, mandated by SUSY.   

\noindent
{\color{blue}{\it An effective model near the SUSY point.}} This two-band model can be extended to finite $\xi$ near the SUSY point, in full agreement with the (kagome) lattice predictions, as promised earlier. In the tight-binding Hamiltonian under arbitrary inhomogeneous fluxes ($\xi\in[0,1]$), the $\xi$-dependence affects only two of the nn bonds in each kagome SUC, with Peierls phases $e^{i[\xi-8(l+1)]\alpha}$ or $e^{i\xi\alpha}$ for the $l$-th SUC within a given MUC. The simplest interpretation of this structure follows by assuming that it represents a linear interpolation between the homogeneous flux case ($\xi=1$) and the SUSY case ($\xi=0$), which implies the identity $e^{i\xi\alpha} = \left[ 1-\xi + \xi e^{i\alpha}\right]$ is to be ascertained for all $\xi$. It is straightforward to show that it holds for small $\xi$ as well as small $\alpha$ (the regime where ALLs exhibit prominent anomalous quantization). This establishes that, for small $\xi$, an effective two-band continuum model given by the Hamiltonian,
\begin{equation}\label{Ham3}
 {\cal F}({\bf q}) = \sum_\mu \left[ b_\mu({\bf q}) + \xi d_\mu({\bf q}) \right] \sigma_\mu\,,
\end{equation}
provides a reliable description to capture the essential physics of an SFB under inhomogeneous fluxes ($b_\mu, d_\mu$ defined in Eq.~\ref{Ham2} and Eq.~\ref{Ham1} respectively). The resulting ALL spectrum follows $E_n \approx -\xi (a_1a_4/8l^2)(1/n)$ for large $n$ \cite{suppl}, linear with the modulation parameter $\xi$ as predicted previously, showing only minor deviation from the earlier results at $\xi=1$ even though the approximated Hamiltonian in Eq.~\ref{Ham3} is valid only close to $\xi=0$.

\noindent
{\color{blue}{\it Transport in ALLs.}} The topological character of the ALLs is accessed through Hall measurements, where the relevant quantity is the energy-resolved Hall conductivity, 
\(
\sigma_H(E) = \sum_n \int_{\rm BZ} d^2{\bf k}\, F_n({\bf k}),
\) 
with $F_n({\bf k})$ denoting the Berry curvature of the $n$-th band and the sum running over all bands below $E$. In lattice calculations, $F_n$ is evaluated on a discretized Brillouin zone for a kagome MUC specified by given $p,q$ adopting the lattice gauge theory approach of Refs.~\cite{fukui2005chern, hatsugai2006topological}. The resulting bands faithfully resemble LLs, being nearly flat and having Berry curvature $F_n$ almost uniform across the BZ with fluctuations ${\cal O}(q^{-1})$. In all cases, and for any value of the inhomogeneity parameter $\xi$ at fixed $p/q$, $\sigma_H$ is observed to converge to integer values, from which the Chern numbers of the individual bands can be unambiguously resolved. It is thereby concluded that each ALL carries Chern number $\pm 1$ \cite{suppl}, reflecting the Fermi–type LL character \cite{gusynin2005unconventional, hatsugai2006topological}.

\noindent
{\color{blue}{\it Discussion.}} We study gapless SFBs under inhomogeneous magnetic fields, both within a continuum framework and lattice models. Our results demonstrate that inhomogeneity provides a means to tune the resultant ALL spectrum, driving a continuous squeezing that leads to the collapse of the entire tower onto zero energy, thereby exemplifying lattice SUSY in a time-reversal symmetry-broken scenario. The SUSY not only illuminates the spectral characteristics of a higher-order band crossing in a perpendicular magnetic field in terms of a square-root HRBS, but also manifests as a lattice analog of the Aharonov-Casher theorem, hosting degenerate zero modes determined solely by the magnetic field, originating from the same operator algebra. Extending the ambit of this theorem to higher-order topological band degeneracies, following our SUSY approach in a continuum setting, poses an exciting open problem for future work. 

Another important next step is to investigate what anomalous quantization arises when flat bands are lifted away from the zero-energy, preserving the chiral symmetry. Since the ALLs are reminiscent of Fermi-type LLs from their topological characterization, their collapse onto a degenerate manifold through flux inhomogeneity is indicative of the appearance of generalized symmetries and algebraic structures in the underlying wavefunctions. When combined with short-range interactions, this parametric control renders novel pathways for exploring the melting of topological order into unconventional charge–density–wave phases \cite{kuroki1992flat, wu2007flat, neupert2011fractional, sun2011nearly, venderbos2012fractional, grushin2015phase, kumar2025origin}. Looking ahead, these predictions appear well within the reach of modern programmable simulators, most notably the optical lattices \cite{cooper2011optical, celi2014syntheticdims, moller2018dice}, and superconducting quantum processors built of flux-tunable transmon qubits with short-range exchange interactions mediated by circuit capacitance \cite{feng2023hofstadterQSim, rosen2024synthetic}, offering a realistic pathway to experimentally realize the proposed physics.  \\

\noindent
{\it Acknowledgment.} SD and KR appreciate stimulating discussions with Kush Saha, Alexander Wietek, Simon Trebst, and Michael J. Lawler. KR acknowledges support from the PMECRG grant (ANRF/ECRG/2024/006260/PMS) from ANRF, India.

\bibliographystyle{apsrev4-1}
\bibliography{ref}

\clearpage

\begin{widetext}

\newcommand{\beginsupplement}{%
  \setcounter{equation}{0}%
  \setcounter{figure}{0}%
  \setcounter{table}{0}%
  \renewcommand{\theequation}{S\arabic{equation}}%
  \renewcommand{\thefigure}{S\arabic{figure}}%
  \renewcommand{\thetable}{S\arabic{table}}%
}

\newcommand{\Eq}[1]{Eq.~(\ref{#1})}
\newcommand{\Fig}[1]{Fig.~\ref{#1}}
\newcommand{\Sec}[1]{Sec.~\ref{#1}}
\newcommand{\Tab}[1]{Table~\ref{#1}}

\begin{center}
    {\bfseries Supplemental Material for\\[0.5ex]
    ``Anomalous Landau Levels in Inhomogeneous Fluxes and Emergent Supersymmetry''}

    \vspace{2ex}

    {Soujanya Datta\textsuperscript{1,2}, Krishanu Roychowdhury\textsuperscript{1,2,3}}

    \vspace{1ex}

    \textsuperscript{1}Theory Division, Saha Institute of Nuclear Physics, 1/AF Bidhannagar, Kolkata 700064, India\\
    \textsuperscript{2}Homi Bhabha National Institute, Training School Complex, Anushaktinagar, Mumbai 400094, India\\
    \textsuperscript{3}Max-Planck-Institut f\"{u}r Physik komplexer Systeme, N\"{o}thnitzer Strasse 38, 01187 Dresden, Germany

    \vspace{3ex}
\end{center}

\beginsupplement

\section{Anomalous Landau levels in homogeneous magnetic field: direct solution of the continuum model}

{\color{black}{

The eigenvalue equation corresponding to the generic two-band Hamiltonian (Eq.~\ref{Ham1} in the main text) in Landau gauge is
\begin{align}
    \begin{pmatrix}
        {\cal H}_{11} & {\cal H}_{12}\\
        {\cal H}_{21} & {\cal H}_{22}
    \end{pmatrix}
    \begin{pmatrix}
        \phi_A(x)\\
        \phi_B(x)
    \end{pmatrix} 
=E
    \begin{pmatrix}
        \phi_A(x)\\
        \phi_B(x)
    \end{pmatrix}, 
\end{align}
where 
\begin{align} 
 {\cal H}_{11} &= -a_1^2 \partial_x^2 + \left(a_3^2 + a_2^2 \right) \left(k + \frac{x}{l^2}\right)^2 - 2ia_1 a_2 \left[\left(k + \frac{x}{l^2}\right)\partial_x + \frac{1}{2l^2}\right] \nonumber \\
 {\cal H}_{12} &= a_4 \left(a_3 - i a_2 \right) \left(k + \frac{x}{l^2}\right)^2 - a_1 a_4 \left[\left(k + \frac{x}{l^2}\right)\partial_x + \frac{1}{2l^2}\right] \nonumber \\
 {\cal H}_{21} &= a_4 \left(a_3 + i a_2\right)\left(k + \frac{x}{l^2}\right)^2 + a_1 a_4 \left[\left(k + \frac{x}{l^2}\right)\partial_x + \frac{1}{2l^2}\right] \nonumber \\
 {\cal H}_{22} &= a_4^2 \left(k + \frac{x}{l^2}\right)^2. 
\end{align}
Decoupling of the two spinor components yields a second-order differential equation for the component $\phi_A$:
\begin{align}
    ({\cal H}_{11} - E)\phi_A(x) - {\cal H}_{12} ({\cal H}_{22} - E)^{-1} {\cal H}_{21} \phi_A(x) = 0
\end{align}
As the region of interest for the ALLs is around $E=0$ ($E<0$), we find that up to the leading order in $E$ [${\cal O}(E^1)$], the effective equation to solve for $\phi_A$ reads  
\begin{align}
    v^2\partial_v^2 \phi_A - 2v \partial_v \phi_A - \left(\frac{7}{4} + \left(\frac{3 a_4^2}{4 E l^4} 
    + \frac{2 a_3}{ a_1 l^2} \right)v^2 + \frac{a_2^2 + a_3^2 + a_4^2}{ a_1^2 l^4 }  v^4\right)\phi_A = 0\,,
\end{align}
substituted $v = x+k l^2$. A further substitution $\nu=v^2$ produces Eq.~\ref{diffeq1} in the main text that has solutions in terms of the confluent hypergeometric functions ($U$) and generalized Laguerre polynomials ($L$) as 
\begin{align*}
    \phi_A &= C_1 e^{\frac{1}{4} \left(\left(\sqrt{16 P+9}+3\right) \log (\nu)-4 \sqrt{R} \nu\right)} U\left(-\frac{-2 Q-\sqrt{16 P+9} \sqrt{R}-2 \sqrt{R}}{4 \sqrt{R}},\frac{1}{2} \sqrt{16 P+9}+1,2 \sqrt{R} \nu\right)\\
    &+ C_2 e^{\frac{1}{4} \left(\left(\sqrt{16 P+9}+3\right) \log (\nu)-4 \sqrt{R} \nu\right)} L\left( {\frac{-2 Q-\sqrt{16 P+9} \sqrt{R}-2 \sqrt{R}} {4 \sqrt{R}}}, {\frac{1}{2} \sqrt{16 P+9}},2 \sqrt{R} \nu\right),
\end{align*}
where $P={7}/{16}$, $Q=3a_4^2/\left(16 El^4\right) + a_3/\left(2 a_1l^2\right)$, $R=\left(a_2^2 + a_{3}^2 + a_{4}^2\right)/\left(4 a_1^2 l^4\right)$. For these solutions to represent quantum wavefunctions in $L^2({\mathbb C})$, {\it i.e.}, to admit square-integrability in the region $[0,\infty)$, the first argument of the special functions above has to take negative integer values, giving rise to the energy quantization noted in Eq.~\ref{enladder1} of the main text. The other spinor component does not belong to $L^2(\mathbb C)$, hence, is discarded.  \\
}}

\noindent
\textbf{Weakly singular limit:} The setting $a_2 = 0,a_3 = a_1, a_4 \approx a_1 d_m$ characterizes the weakly singular limit in which 
\begin{align}
    E_n \approx - \frac{1}{l^2(n+2)} \left( \frac{ 3 a_1^2 d_{max}^2}{16} \right) \approx - \frac{ 3 a_1^2 d_{max}^2}{16l^2} \cdot \frac{1}{n} ~~(\text{large } n)
\end{align}
\textbf{Strongly singular limit:} As $d_{max} \approx 1$ in this limit, which implies $a_3 \approx 0$, the scaling simplifies to (with $a_2=0$)
\begin{align}
    E_n \approx - \frac{1}{l^2(2n+3)}\left | \frac{3  a_1 a_4}{8}\right | \approx - \frac{ 3 a_1 a_4}{16l^2 } \cdot \frac{1}{n} ~~(\text{large } n)
\end{align}

\section{A kagome lattice tight-binding model in inhomogeneous fluxes}

To understand the tight-binding model of spinless electrons on a kagome lattice in inhomogeneous fluxes, we first set a reference, which is the homogeneous limit. In this, each kagome triangle encloses a flux of $\alpha$ while each hexagonal plaquette encloses $6\alpha$, resulting in a net flux of $8\alpha$ in a structural unit cell (SUC). For a magnetic field of strength $B$, $\alpha = B\times {\textrm{area of the triangle}}$. As mentioned in the main text, we set the flux of an SUC to be a rational fraction modulo the flux quantum $\Phi_0$ as $8\alpha/\Phi_0=p/q$, which enlarges the SUC to a magnetic unit cell (MUC) along the direction of the lattice vector ${\bf a}_1=2a(1,0)$, $a$ is the nearest-neighbor distance, the other lattice vector being ${\bf a}_2=2a(1/2, \sqrt{3}/2)$. 

Let us define the position of an arbitrary SUC by the position vector ${\bf R}(m,n) = m {\bf a}_1 + n {\bf a}_2$, where $m,n$ are integers. As the SUC has three inequivalent sites, denoted $A, B$ and $C$, we assign the electronic annihilation operators to these sites respectively as $A_{m,n}, B_{m,n}, C_{m,n}$. The tight-binding Hamiltonian in the absence of any magnetic field is therefore given by
\begin{equation}
{\cal H}_{\rm TB} = t \sum_{ m,n } \Big[ 
\big( B^{\dagger}_{mn} A_{mn} + B^{\dagger}_{mn} A_{m+1,n} \big) 
+ \big( C^{\dagger}_{mn} A_{mn} + C^{\dagger}_{mn} A_{m,n+1} \big) 
+ \big( C^{\dagger}_{mn} B_{mn} + C^{\dagger}_{mn} B_{m-1,n+1} \big) 
+ {\rm H.C.} \Big],
\label{HamSuppl1}
\end{equation}
which features a flat band at the bottom of the spectrum touching a quadratically dispersing band at the $\Gamma$ point of the Brillouin zone (BZ). In the presence of a perpendicular homogeneous magnetic field, the hopping parameter $t$ acquires a complex phase through the Peierls substitution, which is the integral over the vector potential along the hopping path,
\begin{align*}
    t \rightarrow t e^{-\frac{2 \pi i}{\phi_0}\int_{r_i}^{r_j}  {\bf A(r).dr}} = t e^{i \phi_{ij}}.
\end{align*} 
Introducing an additional index $l$ to denote the SUCs within the enlarged MUC (q times the SUC due to the flux content of each SUC being $8\alpha/\Phi_0=p/q$), the tight-binding Hamiltonian modifies to ${\cal H} = {\cal H} _{\rm BA} + {\cal H} _{\rm CA} + {\cal H} _{\rm CB}$, where ${\cal H} _{\rm BA}$ is the total Hamiltonian for the B-A type bonds, and so on. Explicitly,
\begin{align}\label{HamSuppl2}
 {\cal H} _{\rm BA} & = t \sum_{m,n} \Big[ \sum_{l=1}^{q}
 B^{\dagger}_{m,n,l} A_{m,n,l} + \sum_{l=1}^{q-1} B^{\dagger}_{m,n,l} A_{m,n,l+1} + B^{\dagger}_{m,n,q} A_{m+1,n,1} + {\rm H.C.} \Big] \nonumber \\
 {\cal H} _{\rm CA} & = t \sum_{m,n} \Big[ \sum_{l=1}^{q}
 C^{\dagger}_{m,n,l} A_{m,n,l} + \sum_{l=1}^{q} C^{\dagger}_{m,n,l} A_{m,n+1,l} e^{-2 \pi i (8l\alpha)} + {\rm H.C.} \Big] \nonumber \\
 {\cal H} _{\rm CB} & = t \sum_{m,n} \Big[ \sum_{l=1}^{q}
 C^{\dagger}_{m,n,l} B_{m,n,l} e^{2\pi i \alpha} + C^{\dagger}_{m,n,1} B_{m-1,n+1,q} e^{-2\pi i  (8q+7 )\alpha} + \sum_{l=2}^{q}C^{\dagger}_{m,n,l} B_{m,n+1,l-1} e^{2\pi i  (1 - 8l)\alpha}
 + {\rm H.C.} \Big].
\end{align}
Invoking the translation symmetries of the MUC and performing a Fourier transformation $ X_{\bf{R}} =\frac{1}{\sqrt{N}} \sum_{\bf{k}} e^{i\bf{k}\cdot\bf{R}} X_{\bf{k}}$ ($X=A,B,C$), where $N$ is the number of $\bf k$ points in the BZ, we obtain the $(3q \times 3q)$ Bloch Hamiltonian. Diagonalizing this Hamiltonian at the $\Gamma$ point yields the kagome HRBS shown in the main text. For a fixed $q$ and $p \in [1, 8q-1]$, the system exhibits a self-similar fractal energy spectrum that can be manipulated by flux inhomogeneity, {\it i.e.}, by varying the flux through different simplices within a single SUC.
\begin{figure*}[h]
 \centering \includegraphics[width=1\columnwidth]{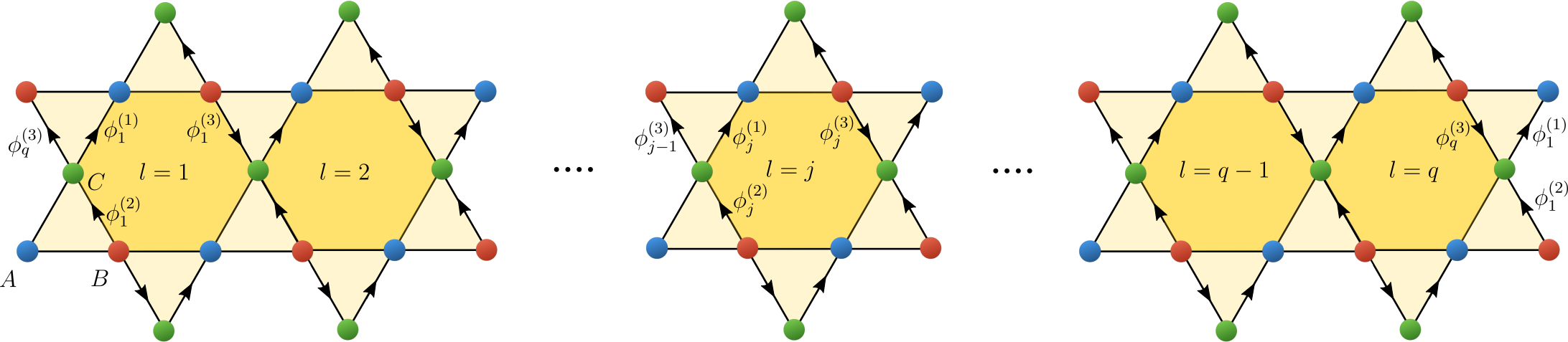}
  \caption{The kagome MUC formed under the perpendicular flux of strength $p/q$ per SUC. The non-zero gauge fields are marked by arrows, and the fluxes through various simplices are mentioned in the text. The shades schematically represent different flux densities for a given inhomogeneous flux configuration.}
 \label{suppfig1}
\end{figure*}

The flux inhomogeneity is introduced by a parameter $\xi$, such that it enables a redistribution of fluxes among the triangles and the hexagon, keeping the total flux content fixed to $8\alpha$. Specifically, the modulation parameter $\xi$ modifies the flux of each triangle $\Phi_\Delta=\alpha \rightarrow \xi \alpha$, while for the hexagon: $\Phi_{\hexagon}=6\alpha \rightarrow (8-2\xi) \alpha$. The homogeneous limit is given by $\xi=1$, and we vary $\xi$ from $1$ to $0$ to explore the inhomogeneous flux configurations relevant to us. Following the convention mentioned in the main text, the resulting tight-binding Hamiltonian for an arbitrary $\xi$ is ${\cal H}(\xi) = {\cal H} _{\rm BA}(\xi) + {\cal H} _{\rm CA}(\xi) + {\cal H} _{\rm CB}(\xi)$ with
\begin{align}\label{HamSuppl3}
 {\cal H} _{\rm BA}(\xi) & = t \sum_{m,n} \Big[ \sum_{l=1}^{q}
 B^{\dagger}_{m,n,l} A_{m,n,l} + \sum_{l=1}^{q-1} B^{\dagger}_{m,n,l} A_{m,n,l+1} + B^{\dagger}_{m,n,q} A_{m+1,n,1} + {\rm H.c.} \Big] \nonumber \\
 {\cal H} _{\rm CA}(\xi) & = t \sum_{m,n} \Big[ \sum_{l=1}^{q}
 C^{\dagger}_{m,n,l} A_{m,n,l} + \sum_{l=1}^{q} C^{\dagger}_{m,n,l} A_{m,n+1,l} e^{-2 \pi i \phi^{(1)}_{l}} + {\rm H.c.} \Big] \nonumber \\
 {\cal H} _{\rm CB}(\xi) & = t \sum_{m,n} \Big[ \sum_{l=1}^{q}
 C^{\dagger}_{m,n,l} B_{m,n,l} e^{2\pi i \phi^{(2)}_{l}} + C^{\dagger}_{m,n,1} B_{m-1,n+1,q} e^{2\pi i \phi^{(3)}_{q}} + \sum_{l=2}^{q}C^{\dagger}_{m,n,l} B_{m,n+1,l-1} e^{2\pi i \phi^{(3)}_{l-1}}  + {\rm H.c.} \Big],
\end{align}
where the $\xi$-dependent phases for the $l$-th SUC are $\phi^{(1)}_l = 8l\alpha, \phi^{(2)}_l = \xi \alpha, \phi^{(3)}_l = [\xi - 8(l+1)]\alpha$ to ensure that each triangular plaquette carries flux $\Phi_{\Delta} = \phi^{(2)}_l = \xi \alpha$, while each hexagon carries flux $\Phi_{\hexagon} = -\phi^{(1)}_l - \phi^{(2)}_l - \phi^{(3)}_l = (8 - 2\xi)\alpha$, independent of $l$ (cf. the main text and Fig.~\ref{suppfig1}). The remainder of the analysis parallels the homogeneous case.

\section{Density of states and edge locality marker}
We compute the density of states (DOS) $\rho(E)$ at a given energy $E$ using a Lorentzian approximation 
\begin{align}
  \rho(E) =  \frac{1}{N} \sum_{n} \frac{\sigma/\pi}{\left(E - E_n\right)^2 + \sigma^2 },
\end{align}
the sum of $n$ running over $N$ bands, and $\sigma$ is a small smoothening parameter. For a given flux of the SUC $\Phi_{\rm SUC}/\Phi_0 \equiv 8\alpha/\Phi_0=p/q$, the MUC supports $N=3q$ bands to sum over. The DOS displayed in the main text is for $p=1, q=113$, and $\sigma=0.025$.

In order to verify the existence of edge states in the major gaps of the HRBS, complementary to the DOS (high bulk DOS implies low edge state density in the spectrum), we evaluate an eigenstate-resolved quantity, called the edge locality marker (ELM) \cite{tran2015topological}, under open boundary conditions, defined as
\begin{align}
    {\cal L}_n = \frac{1}{N_e} \sum_{i \in {\rm edge}} \left| \psi_n (i) \right|^2
\end{align}
where $\psi_n(i)$ is the contribution from the $i$-th edge site for the $n$-th (normalized) eigenvector of the tight-binding Hamiltonian for the system with $L_x=1, L_y=37$ MUCs included, $N_e$ is the number of edge sites. For states characterized by dense DOS, the weight $|\psi_n(i)|^2$ spreads in the bulk, so ${\cal L}_n \approx 0$. On the other hand, the edge marker becomes significant inside the major spectral gaps ({\it i.e.}, regions of low DOS), revealing the population of edge states at those energies. The ELM displayed in Fig.~\ref{suppfig2} is for $p=1, q=37$.
\begin{figure*}
 \centering \includegraphics[width=1\columnwidth]{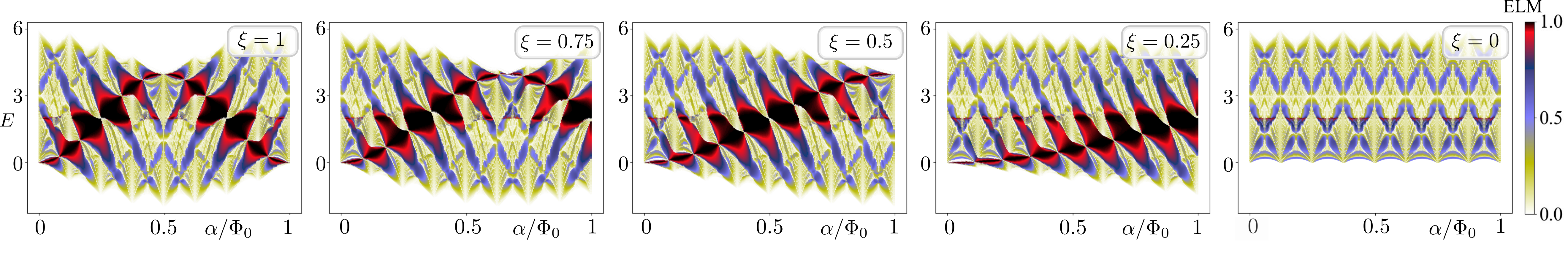}
  \caption{Plot of the edge locality marker ${\cal L}_n$, defined in the text, for the kagome lattice model subjected to inhomogeneous fluxes controlled by the parameter $\xi$. A large value of this marker signifies the major gaps in the HRBS and the existence of sub-gap edge states (other details are in the text). }
 \label{suppfig2}
\end{figure*}

\section{Energy quantization of the anomalous Landau levels in inhomogeneous fluxes}

This section is dedicated to explaining the anomalous quantization of the LLs for an SFB exposed to perpendicular magnetic fields with both a homogeneous pattern and inhomogeneities considered in the main text. We will start with the homogeneous flux limit and later generalize to the case of inhomogeneous fluxes. We will follow Ref.~\cite{rhim2020quantum} for the derivation of the homogeneous case.  \\

\noindent
\textbf{Homogenuous fluxes:}
The generic two-band Hamiltonian, as given in Eq.~\ref{Ham1} of the main text, denoted ${\cal H}(k)$ here, can be expressed in terms of ladder operators in the presence of a uniform magnetic field through the replacement, 
\begin{align}
    k_x &= \frac{1}{\sqrt{2}\,l}\,\big(a + a^{\dagger}\big), \qquad
    k_y = \frac{i}{\sqrt{2}\,l}\,\big(a - a^{\dagger}\big), \qquad k_x k_y =
    \frac{1}{2}\big(k_x k_y + k_y k_x\big) = \frac{i}{2l^{2}}\,
    \big(a^{2} - a^{\dagger 2}\big),
\end{align}
where $a$ ($a^\dagger$) is the annihilation (creation) operator, and the symmetrization of $k_x, k_y$ is made to ensure the Hermiticity of ${\cal H}$. The bosonic algebra $[a,a^{\dagger}] = 1$ derives from $[k_x,k_y] = - i/l^2$. Following the above replacement, each element of the Hamiltonian matrix has the form
\begin{align}
    {\cal H}_{11} &= \frac{1}{2 l^2} \Big[ 
      \big(a_1^2 - a_3^2 - a_2^2 + 2 i a_1 a_2\big) a^2 
      + \big(a_1^2 - a_3^2 - a_2^2 - 2 i a_1 a_2\big) a^{\dagger 2}  \nonumber  + \big(a_1^2 + a_3^2 + a_2^2\big)(2 a^\dagger a + 1) \Big], \\[6pt]
    {\cal H}_{12} &= \frac{1}{2 l^2} \Big[
      a_4(a_1 - a_3 + i a_2)\, a^2
      + a_4(i a_2 - a_1 - a_3)\, a^{\dagger 2} 
      + a_4(a_3 - i a_2)(2 a^\dagger a + 1) \Big],\\[6pt]
    {\cal H}_{21} &= H_{12}^\dagger,\\[6pt]
    {\cal H}_{22} &= - \frac{a_4^2}{2 l^2} 
      \Big(a^2 + a^{\dagger 2} - 2 a^\dagger a - 1 \Big).
\end{align}
To obtain the spectrum, we express the above Hamiltonian in the basis of the number operator $\hat{n}=a^\dagger a$ and solve the eigenvalue equation using a trial wavefunction in this basis as
\begin{align}
    |\Psi \rangle = \sum_{n=0}^{\infty} 
\begin{pmatrix}
           A_n \\
           B_n 
\end{pmatrix} 
|n \rangle\,.
\end{align}
For computation, we truncate the basis at some $n=N_c$ to obtain a finite-dimensional representation of ${\cal H}$ which is now a $(2N_c \times 2N_c)$ matrix of the form
\begin{align}
    {\cal H}_{LL} = \frac{1}{l^2}
    \begin{pmatrix}
        P_0 & 0 & Q_0 & 0 & 0 &.&.&.\\
        0 & P_1 & 0 & Q_1 & 0 & .&.&.\\
        Q_0^\dagger & 0 & P_2 & 0 & Q_2 & .&.&. \\
        0 & Q_1^\dagger & 0 & P_3 & 0 & .&.&.\\
        . & . & . & . & .&.&.&. \\
        . & . & . & . & .&.&.&. \\
        . & . & . & . & . &.&.&.
    \end{pmatrix},
\end{align}
(the subscript $LL$ to denote its diagonalization giving rise to LLs) in the basis $( A_1, B_1, ... , A_{N_c}, B_{N_c})^T$, where
 \begin{align}
     P_n &= (2n+1)
\begin{pmatrix}
a_3^2 + a_1^2 + a_2^2 & a_4 (a_3 - i a_2)\\
a_4 (a_3 + i a_2) &  a_4^2    
\end{pmatrix} ~~;~~
Q_n = \sqrt{(n+1)(n+2)}
\begin{pmatrix}
  a_1^2 - a_3^2 - a_2^2 + 2 i a_1 a_2 & a_4 (a_1 - a_3 + i a_2)\\
-a_4 (a_1 + a_3 + i a_2) &  -a_4^2 
\end{pmatrix}
\end{align}
The form of ${\cal H}_{LL}$ is immediately suggestive of the coupling to be nonzero only between blocks $n$ and $n\pm 2$, so that for a given $n$, the following eigenvalue equation is to be solved
\begin{align}
     \frac{1}{2 l_{B}^2}
     \begin{pmatrix}
         P_n & Q_n\\
         Q_{n}^\dagger & P_{n+2}
     \end{pmatrix}
          \begin{pmatrix}
         \phi_n \\
         \phi_{n+2}
     \end{pmatrix}
     = E 
     \begin{pmatrix}
         \phi_n \\
         \phi_{n+2}
     \end{pmatrix}.
\end{align}
Projecting onto the subspace of the block indexed by $n$, this yields an effective ($2 \times 2$) Hamiltonian for the LLs 
\begin{align}
    {\cal H}_{LL}^n(E) = \frac{1}{2 l^2} [P_n + Q_n(E-P_{n+2})^{-1}Q_{n}^\dagger]\,.
\end{align}
As the ALLs appear at energies very close to $E=0$, a leading order expansion in $E$ gives rise to the effective LL Hamiltonian
\begin{align}
   {\cal H}_{LL}^n(0) &\approx \frac{1}{2 l^2} [P_n + Q_n(-P_{n+2})^{-1}Q_{n}^\dagger],
\end{align}
which in the large $n$ limit, has the spectrum 
\begin{align}
    E_n = -\frac{3 a_1^2 a_4^2}{(2n+5)(a_3^2 + a_1^2 + a_2^2 + a_4^2 + 2a_3 a_1)l^2}\,.
\end{align}
\textbf{Weakly singular limit:} The setting $a_2 = 0,a_3 = a_1, a_4 \approx a_1 d_{max}$ characterizes the weakly singular limit in which 
\begin{align}
    E_n \approx - \frac{3a_1^2 d_{max}^2}{8l^2} \cdot \frac{1}{n}~~(\text{large } n)\,.
\end{align}
\textbf{Strongly singular limit:} As $d_{max} \approx 1$ which implies $a_3 \approx 0$, the scaling simplifies to (with $a_2=0$)
\begin{align}
    E_n \approx - \frac{3a_1^2 a_4^2}{2(a_1^2 + a_4^2)l^2}\cdot \frac{1}{n} ~~(\text{large } n)\,.
\end{align} \\

\noindent 
\textbf{ALLs at the SUSY point (continuum):}
The continuum model for realizing ALLs at the SUSY point can be best understood by squaring the continuum form of the charge operator $\cal Q$ in Eq.~\ref{charge} of the main text, then introducing the ladder operators as before, to obtain a ($2\times 2$) Hamiltonian ${\cal G}$ with elements 
\begin{align}
    {\cal G}_{11} &= \frac{1}{2 l^2} \Big[ 
      \big(a_1^2 - a_3^2 - a_2^2 + 2 i a_1 a_2\big) a^2 
      + \big(a_1^2 - a_3^2 - a_2^2 - 2 i a_1 a_2\big) a^{\dagger 2}  \nonumber  + \big(a_1^2 + a_3^2 + a_2^2\big)(2 a^\dagger a + 1) -2a_1 a_3 \Big], \\[6pt]
    {\cal G}_{12} &= \frac{1}{2 l^2} \Big[
      a_4(a_1 - a_3 + i a_2)\, a^2
      + a_4(i a_2 - a_1 - a_3)\, a^{\dagger 2} 
      + a_4(a_3 - i a_2)(2 a^\dagger a + 1)-a_1 a_4 \Big],\\[6pt]
    {\cal G}_{21} &= {\cal G}_{12}^\dagger,\\[6pt]
    {\cal G}_{22} &= - \frac{a_4^2}{2 l^2} 
      \Big(a^2 + a^{\dagger 2} - 2 a^\dagger a - 1 \Big).
\end{align}
With the above procedure repeated, its matrix form in the number operator basis reads
\begin{align}
    {\cal G}_{LL} = \frac{1}{2 l^2}
    \begin{pmatrix}
        P'_0 & 0 & Q'_0 & 0 & 0 &.&.&.\\
        0 & P'_1 & 0 & Q'_1 & 0 & .&.&.\\
        Q_0'^{\dagger} & 0 & P'_2 & 0 & Q'_2 & .&.&. \\
        0 & Q_1'^{\dagger} & 0 & P'_3 & 0 & .&.&.\\
        . & . & . & . & .&.&.&. \\
        . & . & . & . & .&.&.&. \\
        . & . & . & . & . &.&.&.
    \end{pmatrix},
\end{align}
where 
 \begin{align}
     P'_n &=
\begin{pmatrix}
 (2n+1)(a_3^2 + a_1^2 + a_2^2) - 2 a_3 a_1 &  (2n+1)a_4 (a_3 - i a_2) - a_1 a_4\\
 (2n+1)a_4 (a_3 + i a_2) - a_1 a_4 &   (2n+1)a_4^2    
\end{pmatrix} \nonumber \\ Q'_n &= \sqrt{(n+1)(n+2)}
\begin{pmatrix}
  a_1^2 - a_3^2 - a_2^2 + 2 i a_1 a_2 & a_4 (a_1 - a_3 + i a_2)\\
-a_4 (a_1 + a_3 + i a_2) &  -a_4^2    
\end{pmatrix}
 \end{align}
Comparing with the homogeneous case, the off-diagonal blocks remain the same, $Q_n = Q'_n$, whereas the diagonal blocks differ due to an additional term, $P_n \neq P'_n$.
Solving in the same manner as before, we obtain the spectrum
\begin{align}
    E_n = 0, \frac{n}{2 l_{B}^2}(a_3^2 + a_1^2 + a_2^2 + a_4^2 + 2 a_3 a_1)\,.
\end{align}
\\

\noindent
\textbf{Intermediate inomogenuous fluxes:} The inhomogeneity of fluxes is introduced via the parameter $\xi$, interpolating linearly between the homogeneous limit and the SUSY limit. Drawing insights from the variation of the spectrum in terms of $\xi$ (Fig.~\ref{fig3}) in our lattice model, we consider the continuum Hamiltonian for an arbitrary $\xi$ as a linear interpolation 
\begin{align*}
    {\cal F}_{LL}(\xi) = \xi {\cal H}_{LL} + (1-\xi){\cal G}_{LL}
\end{align*}
This effective model successfully captures the influence of inhomogeneity in close agreement with the lattice results, with only the matrix $P_n$ modified to
\begin{align}
{\tilde{P}}_n =
\begin{pmatrix}
(1+2n)\,(a_1^{2}+a_2^{2} + a_3^2) + 2(\xi - 1)\,a_3a_1
&
\Big[ (1+2n)\,(a_3 - ia_2) + (\xi - 1)\,a_1  \Big] a_4
\\[6pt]
\Big[ (1+2n)\,(a_3 + i a_2) + (\xi - 1)\,a_1\Big] a_4
&
(1+2n)\,a_4^{2}
\end{pmatrix}
\end{align}
while $Q_n$ remains the same, {\it i.e.}, $\tilde{Q}_n = Q_n$. We have chosen $a_2 = a_3 = 0$ for simplification. The effective intermediate Hamiltonian finaly takes the form
\begin{align}
   {\cal F}_{n}(\xi) =  \left(
\begin{array}{cc}
 \frac{a_1^2 (\xi +2 n (\xi +n+2))}{2 l^2 (\xi +2 n+4)} & \frac{a_1 a_4 \left(\xi  (\xi +3)+2 n^2+2 (\xi +2) n\right)}{2 l^2 (\xi +2 n+4)} \\
 \frac{a_1 a_4 \left(\xi  (\xi +3)+2 n^2+2 (\xi +2) n\right)}{2 l^2 (\xi +2 n+4)} & \frac{a_4^2 (\xi +2 n (\xi +n+2))}{2 l^2 (\xi +2 n+4)} \\
\end{array}
\right),
\end{align}
whose spectrum is
\begin{align*}
E^{\pm}_{n}\ =    \frac{\left(\xi + 2n\left(\xi + n + 2\right)\right)\left(a_1^2+a_4^2\right) \pm \sqrt{4 a_1^2 a_4^2 \xi  (\xi +2) (2 n+\xi ) (2 n+\xi +4)+\left(\xi + 2n\left(\xi + n + 2\right)\right)^2 \left(a_1^2+a_4^2\right){}^2 }}{4 (2 n+\xi +4)}
\end{align*}
Focusing on the regime close to zero energy, we consider $E^{-}_{n}$
, and in the vicinity of the SUSY limit $(\xi \rightarrow0)$ with large $n$, the leading-order term reveals the quantization
\begin{align}
 E^{-}_{n}\approx   -\xi\frac{a_1^2 a_4^2}{\left(a_1^2+a_4^2 \right) l^2 } \cdot \frac{1}{n}~~(\text{large } n)\,.
\end{align}
In the main text, it has been argued that this is indeed the description of an SFB in inhomogeneous fluxes close to the SUSY point ($\xi=0$). 

\begin{figure*}
 \centering \includegraphics[width=0.6\columnwidth]{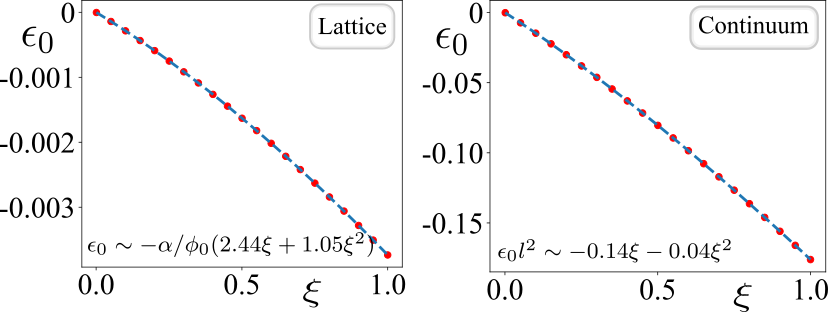}
  \caption{{\bf Left:} Plot of the lowest ALL energy $\epsilon_0$ as a function of the modulation parameter $\xi$ for the kagome MUC with specifications $p=1,q=113,\alpha = p/8q$. Small variations in the coefficients of $\xi$ and $\xi^2$ are observed with $q$. {\bf Right:} The same obtained from continuum calculations by solving the truncated Hamiltonian ($2N_c \times 2N_c$) with $N_c=100$, $a_1 = 0.7, a_2 =1, a_3 =0.9, a_4 =1$. We find the coefficients of $\xi$ and $\xi^2$ have insignificant variation with $N_c$, and the linear term suffices to produce a faithful description.}
 \label{fig3}
\end{figure*}

\section{Direct solution of the effective Hamiltonian near the SUSY point}
An effective theory of SFBs under inhomogeneous fluxes in the vicinity of the SUSY point can be derived by treating $\xi$ as a perturbative parameter multiplied by the Hamiltonian in the homogeneous limit. Added to the SUSY Hamiltonian ${\cal G}({\bf k})$, the full Hamiltonian at an arbitrary small value of $\xi$ is taken to be (cf. Eq.~\ref{Ham3} of the main text) 
\begin{align*}
   {\cal F}({\bf k},\xi) = {\cal G}({\bf k}) +\xi{\cal H}({\bf k})
\end{align*}
For simplification, we choose $a_2 = a_3 = 0$. Using the same gauge for minimal substitution, as in the main text, we obtain the confluent hypergeometric differential equation to solve
\begin{align}
    \nu^2\partial_\nu^2 \phi_A - \frac{\nu}{2} \partial_\nu \phi_A - [L + M\nu + N  \nu^2]\phi_A = 0\,,
\end{align}
where $\nu = \left(x + kl^2 \right)^2$, $L={\xi(7\xi+6)}/{16(1+\xi)^2}$, $M = a_4^2 \xi (3\xi + 2)/\left(16 El^4(1+\xi)\right) $, $N= a_{4}^2/\left(4 a_1^2 l^4\right)$. The general solutions are of the form
\begin{align*}
    \phi_A &= C_1 e^{\frac{1}{4} \left(\left(\sqrt{16 L+9}+3\right) \log (\nu)-4 \sqrt{N} \nu\right)} U\left(-\frac{-2 M-\sqrt{16 L+9} \sqrt{N}-2 \sqrt{N}}{4 \sqrt{N}},\frac{1}{2} \sqrt{16 L+9}+1,2 \sqrt{N} \nu\right)\\
    &+ C_2 e^{\frac{1}{4} \left(\left(\sqrt{16 L+9}+3\right) \log (\nu)-4 \sqrt{N} \nu\right)} L\left( {\frac{-2 M -\sqrt{16 P+9} \sqrt{N}-2 \sqrt{N}} {4 \sqrt{N}}}, {\frac{1}{2} \sqrt{16 L+9}},2 \sqrt{N} \nu\right),
\end{align*}
and the requirement of square integrability leads to the energy quantization
\begin{align}
    \frac{-2 M-\sqrt{16 L+9} \sqrt{N}-2 \sqrt{N}}{4 \sqrt{N}} = -n\,.
\end{align}
Substituting the expressions of $N,M$ and $L$,
\begin{align*}
     \frac{1}{16} \left(\frac{a_1 a_4 \xi  (3 \xi +2)}{E l^2 (\xi +1)}+\frac{4(4 \xi +3)}{(\xi +1)}+8\right) = -n\,,
\end{align*}
and inverting the solution (of the quadratic equation of $\xi$), we obtain
\begin{align}
    E_n = -\frac{a_1 a_4 \xi  (3\xi +2)}{4 l^2 \left(4 n+5 + 6\xi+4\xi n \right)}    \approx  -\xi\frac{a_1 a_4 }{8 l^2}\cdot \frac{1}{n}~~(\text{large } n)\,.
\end{align}
\begin{figure*}[h]
 \centering \includegraphics[width=1\columnwidth]{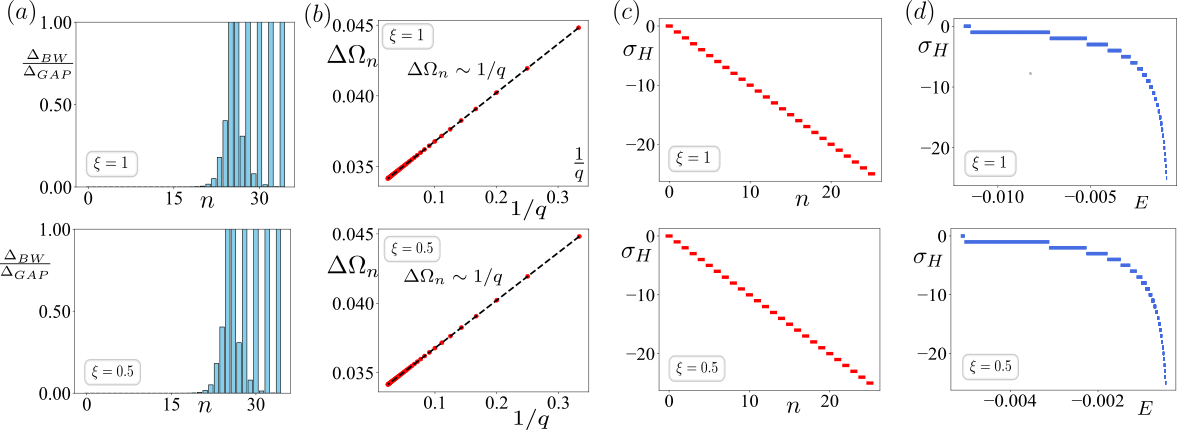}
  \caption{(a) The ratio of $\Delta_{BW}$ to $\Delta_{GAP}$ (defined in the text) plotted against the band index $n$ for the bands below $E=0$. A low value of this ratio is indicative of the nearly flat nature of these bands in close analogy to LLs. (b) The behavior of the quantity $\Delta \Omega$ (defined in the text) with $q$ -- we obtain $\Delta \Omega \sim 1/q$, suggesting, at small enough magnetic fields, the Berry curvature of the lattice eigenstates approaches that of the corresponding LLs. (c) Plot of $\sigma_H$ vs the band index $n$ implies the Chern number of each of these bands is $-1$, similar to the LLs. (d) The same information shown through the plot of $\sigma_H$ vs $E$ reveals the region over which an anomalous energy quantization exists, together signifying these bands to reflect the features of ALLs in lattice models. For all the plots, the top corresponds to the homogeneous case $\xi=1$, and the bottom, a representative inhomogeneous case with $\xi=0.5$. The plots are obtained at $p=1, q=37$.}
 \label{suppfig4}
\end{figure*}

\section{Characterization of the ALLs}

In general, the energy levels of the HRBS are not exact LLs by definition; instead, they are sub-bands formed by rational magnetic flux $p/q$ per SUC. However, in a sufficiently weak magnetic field, they exhibit spectral and topological characteristics similar to LLs, which can be established based on the following observations, summarized in Fig.~\ref{suppfig4}. 

\begin{itemize}
 \item For very small $p$, or large $q$, the ratio of the bandwidth ($\Delta_{BW}$) of the individual bands to the average gap ($\Delta_{GAP}$) with the nearest bands is very low for the lowest few bands, which makes them appear nearly flat.  
 \item The Berry curvature of these bands ($\Omega_n$) admits nonuniformity over the BZ; however, the fluctuations get suppressed with the number of mesh points inside the BZ used to compute $\Omega_n$, as well as with $q$. Specifically, the quantity $\Delta \Omega_n = \langle \Omega_n({\bf {k}}) \rangle /\Omega_{LL} - 1$, where the LL Berry curvature is $\Omega_{LL} = -2\pi/{A_{MBZ}}$, $A_{MBZ}$ being the area of the magnetic Brillouin zone, falls as $1/q$.   
 \item The Hall conductivity $\sigma_H$ calculated following Ref.~\cite{fukui2005chern} and plotted against the ALL index $n$ reveals these ALLs to have Chern number $-1$, which can also be extracted from the plot of $\sigma_H$ vs $E$ to demarcate the region in which the anomalous energy quantization is obtained. The steps of $e^2/h$ are suggestive of their Fermi-type nature. 
\end{itemize}
\end{widetext}

\clearpage
\end{document}